\documentclass[twocolumn]{aa}
\usepackage{graphicx}
\usepackage{natbib}
\bibpunct{(}{)}{;}{a}{}{,} 
\usepackage{txfonts}
\newcommand{\degg}[1]         {${#1}^{\circ}$}

\newcommand{\mum}[1]          {$#1\ \mathrm{\mu m}$}
\newcommand{\opac}[1]         {$#1\ \mathrm{m^{2} kg^{-1}}$}

\newcommand{\anyexp}[3]       {${#1}\times10^{#2}$~{#3}}

\newcommand{\fovp}[2]         {$#1\times#2$}

\newcommand{\msun}[1]         {$#1\ \mathrm{M}_{\odot}$}

\newcommand{\rsuno}           {$\mathrm{R}_{\odot}$}

\newcommand{\aviapp}[1]       {$\mathrm{A}_{\mathrm{v}} \sim #1\ \mathrm{mag}$}
\newcommand{\avio}            {$\mathrm{A}_{\mathrm{v}}$}
\newcommand{\akapp}           {$\mathrm{A}_{\mathrm{K}}$}
\newcommand{\kmsc}            {$\mathrm{kms}^{-1}$}

\newcommand{\molh}            {$\mathrm{H}_{2}$}

\newcommand{\tausi}           {$\mathrm{\tau_{Si}(9.7 \mu m)}$} 

\newcommand{\eng}[2]          {$#1\times10^{#2}$}

\newcommand{\macc}[2]         {$#1\times 10^{#2}\ \mathrm{M_{\odot}yr^{-1}}$}
\newcommand{\macco}           {$\mathrm{M_{\odot}yr^{-1}}$}

\newcommand{\hk}              {$\mathrm{H-K}$}

\newcommand{\rcen}            {$\mathrm{{r}_{c}}$}

\newcommand{\rout}            {$\mathrm{R_{out}}$}
\newcommand{\rcav}            {$\mathrm{R_{cav}}$}
\newcommand{\rd}              {$\mathrm{r_{d}}$}
\newcommand{\rst}             {$\mathrm{R_{*}}$}

\newcommand{\rce}             {$\mathrm{r_{c}}$}

\newcommand{\mac}             {$\mathrm{\dot{M}}$}

\newcommand{\kkext}           {$\mathrm{\kappa^{ext}_\lambda}$}
\newcommand{\kkextH}          {$\mathrm{\kappa^{ext}_H}$}

\newcommand{\od}              {$\mathrm{\tau_\mathrm{los}}$}
\newcommand{\opa}             {$\mathrm{2\theta_{lim}}$}
\newcommand{\floh}            {$\mathrm{f_{H}^{0}}$}
\newcommand{\flok}            {$\mathrm{f_{K}^{0}}$}

\newcommand{\fapprec}         {$\mathrm{F_{app}/F_{rec}}$}

\newcommand{\fout}            {$\mathrm{\mathcal{F}_{out}}$}

\newcommand{\fluunit}[2]      {$\mathrm{erg\ s}^{-1} \mathrm{cm}^{-2}$}

\newcommand{\ras}[1]          {$\mathrm{\alpha\ (h m s)}$}
\newcommand{\dec}[1]          {$\mathrm{\delta\ (^{\circ} {'} {''})}$}

\newcommand{\magarc}[1]       {$\mathrm{#1\ mag/arcsec^2}$}

\newcommand{\mon}             {\object{Mon\,R2\,IRS3}}

\newcommand{\sonef}           {\object{S140\,IRS1}}

\begin{document}
   \title{Constraints in the circumstellar density distribution of
   massive Young Stellar Objects}

   \subtitle{}

   \author{C. Alvarez\inst{1,2,3}
          \and
          M. Hoare\inst{1}
          \and
	  P. Lucas\inst{4}
          }

   \offprints{C. Alvarez, \email{alvarez@mpia-hd.mpg.de}}

   \institute{Physics and Astrononomy Department, University of Leeds,
              Leeds LS2 9JT, United Kingdom\\
              \email{alvarez@mpia-hd.mpg.de}
         \and
              Kapteyn Astronomical Institute, Postbus 800, 9700 AV Groningen, 
              The Netherlands
         \and
	      Max-Planck-Institut f\"ur Astronomie, 
	      K\"onigstuhl 17,  D-69117 Heidelberg,  Germany 
              \and
	      Dept. of Physical Sciences, University of Hertfordshire, 
	      College Lane, Hatfield AL10 9AB, United Kingdom
   }

   \date{Received September 01, 2003; accepted March 01, 2003}

   \abstract{
   We use a Monte Carlo code to generate synthetic near-IR reflection
   nebulae that resemble those (normally associated with a bipolar 
   outflow cavity) seen towards massive young stellar objects (YSOs). 
   The 2D axi-symmetric calculations use an analytic expression for a
   flattened infalling rotating envelope with a bipolar cavity
   representing an outflow. We are interested in which aspects of the
   circumstellar density distribution can be constrained by
   observations of these reflection nebulae.  We therefore keeep
   the line of sight optical depth constant in the model grid, as this
   is often constrained independently by observations.  
   It is found that envelopes with density distributions corresponding to 
   mass infall rates of $\sim 10^{-4} $\macco\ (for an envelope
   radius of 4700 AU) seen at an inclination 
   angle of $\sim $\degg{45} approximately reproduce the morphology and
   extension of the sub-arcsecond nebulae observed in massive YSOs.
   Based on the flux ratio between the approaching and receding lobe of
   the nebula, we can constrain the system inclination angle. The
   cavity opening angle is well constrained from the nebula opening
   angle. Our simulations indicate that to constrain the outflow cavity shape 
   and the degree of flattening in the envelope, near-IR imaging with 
   higher resolution and dynamic range than speckle imaging in
   4m-class telescopes is
   needed. The radiative transfer code is also used to simulate the 
   near-IR sub-arcsecond nebula seen in \mon. We find indications of a
   shallower opacity law in this massive YSO than in the interstellar
   medium, or possibly a sharp drop in the envelope density distribution
   at distances of $\sim$ 1000~AU from the illuminating source.
   }

   \maketitle
%

\section{Introduction}
\label{IntroductionSection}

Bipolar outflows appear to be a ubiquitous phenomenon during the
formation of stars in all mass ranges
\citep{BallyLada83,Henning00,RidgeMoore01,Beuther02}. Low mass young stellar
objects (YSOs) show highly collimated bipolar jets from a few 
10 ~AU \citep{Burrows96} to several parsec
\citep{Reipurth97,Eisloeffel00} in length. These jets are thought to be
magneto-hydrodynamically collimated in a wind formed at the inner
star-disk system \citep[e.g. X-wind, ][]{Shu94}. The jets are thought
to drive the large scale molecular outflow \citep{Masson94}. 

The formation and collimation of outflows in massive YSOs is less well 
understood than in low mass YSOs. There appears to be a lack of
highly collimated parsec-scale jets \citep{Mundt94}. In the near-IR, 
searches for shock-excited \molh\ show traces of jets in massive star 
forming regions, but probably driven by low mass young stars located
in the same cluster \citep{Davis98,Wang03}. A recent search
for optical shock-excited emission in the outer parts of the outflow, 
yielded no evidence of jet interaction (Alvarez \& Hoare, in prep.). Very
close to the driving source, there is also not clear evidence that jets
are the rule in massive YSOs. In some cases, the free-free radio
emission from the inner wind shows a jet morphology (e.g. HH80-81, 
\citealt{Marti93}; Cep~A, \citealt{Torrelles96}). Such jets would have
to be magneto-hydrodynamically driven, even though the OB stars
themselves are not magnetically active. Magneto-hydrodynamics in the
infalling rotating cloud could set up bipolar flows \citep{Tomisaka98}.
In other cases, the ionised wind appears to be
equatorial \citep[e.g.][]{Hoare94,HoareMuxlow96,Hoare02}. Theoretical
models show that radiation pressure in massive young stars can drive
gas off the surface of a disk, producing a predominantly
equatorial wind \citep{Drew98, Drew00}. Any initial flow maybe
hydrodynamically collimated into a bipolar flow by the flattened
surrounding cloud \citep[e.g. ][]{Delamarter00}. These alternative
theories will predict different morphologies for the base of the
outflow cavities carved out. These variations in morphology occur 
at scales of a few 100~AU, which at the typical distances to massive
YSOs of $\sim 1$~kpc, correspond with angular sizes of
$\sim$~0\farcs1. Therefore, high resolution techniquies are
fundamental to study the impact of the the outflow in the surrounding
material.  

In a related paper (Alvarez et al. in prep., hereafter Paper~I), we
show high resolution near-IR speckle images which trace the
circumstellar matter around massive YSOs at 
scales of a few 100~AU. The extended emission that is seen towards
some of the sources can be interpreted as scattered light in an
outflow cavity due to its monopolar morphology. This intepretation is 
supported in some cases by the blue colours of the nebula
\citep[e.g. \mon, Paper I,][]{Preibisch02}. Furthermore, polarimetric 
speckle imaging of the reflection nebula in the massive star forming 
region \sonef\ \citep{Schertl00} shows a centrosymmetric pattern which
is typical of scattered light.

Intuitively, one can imagine that depending on the properties
of the dust, the shape of the cavity, the density distribution and 
the orientation of the system with respect to the observer, the
resulting reflection nebula will change. The morphology of the cavity
is particularly important because it is shaped by the interplay
between the infall and the outflow. For instance, it is expected that an
equatorial or wide-angled wind will produce a cavity with a wide
openig angle near the star. However, a jet is expected to open a
rather narrow cavity.

Radiative transfer simulations have been widely used to generate
synthetic nebulae that resemble the observations
\citep{Lazareff90,WhitneyHartmann92,Kenyon93,Fischer94,Fischer96,Whitney97,Lucas97,Lucas98,Wolf02}. 
The work by \citet{Lazareff90} was based on a ray-tracing code and it was
focused mainly on the effect produced by different disc models on the
synthetic nebulae. The authors compared the general features of the
model images with previous seeing-limited images of the low
mass systems HL~Tau and L\,1551\,IRS5. 
\citet{WhitneyHartmann92}, \citet{Kenyon93} and \citet{Whitney97} developed
a Monte Carlo code to investigate how the nebula morphology and the near-IR
colours of the synthetic images vary with different model parameters.
In particular, \citet{Whitney97} used their code to constrain the colours
of the central source, the dust model and the envelope density
distribution in a sample of $\sim$\,20 low mass
YSOs. \citet{Fischer94} and
\citet{Fischer96} developed a new Monte Carlo scattering code and they
focused on exploring the effect of different dust models in 
the synthetic images. \citet{Lucas97} and \citet{Lucas98}
compared synthetic nebulae produced with their Monte Carlo code with high
resolution multi-colour observations of reflection nebulae associated 
with low mass YSOs. From this comparison, they could constrain
some parameters defining circumstellar density distribution as well as
the dust model for several sources. Recently, radiative transfer 
Monte Carlo codes have been developed to simulate scattering by
non-spherical dust particles \cite{Whitney02,Wolf02,Lucas03}.

These previous models have focused predominantly on low mass
YSOs. Here, we apply the Monte Carlo code of \citet{Lucas98} to high
mass YSOs, where the infall rates are much higher. We also adopt an 
observational approach, by presenting a grid of
models in which as each parameter is varied, the overall density 
is scaled too to keep the optical depth along the
line of sight constant. This is because the line of sight
optical depth is often well constrained from other data such as the
optical depth of the 9.7~$\mu$m silicate feature or the colour of 
the star. The models are decribed in
Sect.~\ref{ModelsSection}. The grid of models is
presented in Sect.~\ref{GridSection}. In
Sect.~\ref{IRS3ModelSection}, we use the models to
constrain the density distribution in
\mon. Some concluding remarks are shown in Sect.~\ref{ConclusionsSection}.

\section{Models}
\label{ModelsSection}
We used the Monte Carlo code of \citet{Lucas97,Lucas98} with a set of 
parameters adapted to massive YSOs. The models consist of a central
star surrounded by a dusty flattened envelope. A disc characterizes
the density distribution near the equator. The model also includes an 
empty bipolar cavity opened by the outflow in the circumstellar
matter. The photons emerging from the central source are scattered off
the dust grains in the envelope and disc. Each photon can suffer
several  scattering processes until it is either absorbed, or escapes. 
All photons traveling in a particular direction are binned, and projected 
onto the image plane.

The envelope is described by a density distribution resulting from
the collapse of a slowly rotating cloud \citep{Ulrich76,Terebey84}.
The density ($\rho$) at any point (r, $\mu$) is given by Eq.~\ref{eq47},  
\begin{equation}
\label{eq47}
   \mathrm{\rho(r,\mu) \sim \frac{\dot{M}}{8\pi r_{c}(GM)^{1/2}} \cdot \frac{1}{(1 + \mu_{0})^{1/2}} \cdot \frac{1}{r^{1/2}}}
\end{equation}
\begin{equation}
\label{eq43}
   \mathrm{\mu_{0}^{3} + (\frac{r}{r_{c}} - 1)\mu_{0} - \frac{r}{r_{c}}\mu = 0}
\end{equation}
where \mac\ represents the mass infall rate and M is the mass of the
central source. \rce\ is the centrifugal radius, which determines the
degree of flattening of the distribution. Flatter density
distributions are characterised by larger centrifugal radii.   
$\mu = \cos(\theta)$, where $\theta$\ is the position angle of each
particle with respect to the polar axis. $\mu_0 = \cos(\theta_0)$\ 
represents the initial position angle of each infalling particle. 
The equation of motion of the infalling particles (Eq.~\ref{eq43}) 
should be satisfied at every point \citep[see ][ for details]{Ulrich76}.

The disc plays a passive role in the models presented here. It absorbs
and scatters the radiation from the central star, but it does not emit. 
We use Eq.~\ref{eq48} to describe the disc density structure,
\begin{equation}
   \label{eq48}
   \mathrm{\rho_{d}(R,z) = \rho_{0}(R/R_{*})^{-\alpha}e^{-\frac{z^{2}}{2H(R)^{2}}}}
\end{equation}
where R and z satisfy the relation $\mathrm{r^{2} = R^{2} + z^{2}}$, 
$\rho_{0}$\ represents the density in the midplane at the surface of 
the star, and $\mathrm{R_{*}}$\ is the radius of the star. 
$\mathrm{H(R) = H_{0}(R/R_{*})^{\beta}}$\ is the disc scale height.
$\beta$\ parametrises the degree of flaring on the disc.  
We tried steady Keplerian discs ($\alpha=15/8$ and $\beta=9/8$) as well as 
geometrically thin and optically thick discs ($\alpha=3/4$ and
$\beta=0$) \citep{Lazareff90,WhitneyHartmann92}.

An empty cavity represents the material evacuated by the outflow in the
envelope. The shape of the cavity must be determined by the interplay 
between the infall and outflow processes. However, the lack of knowledge 
of these processes makes the choice of the cavity shape somewhat arbitrary. 
For some of the models a conical cavity was used,
\begin{equation}
   \label{eq423}
   \mathrm{z = B \cdot (R - R_{cav})}
\end{equation}
where B represents the tangent of the cavity half-opening angle.
For other models, a parabolic cavity was used, which is represented by 
the expression,
\begin{equation}
   \label{eq422}
   \mathrm{z = A \cdot (R^{2} - R_{cav}^{2})}
\end{equation}
where the constant A determines the curvature of the parabola, and 
\rcav\ represents the radius of the cavity at the equator. The 
cavity opening angle (\opa) is defined as twice the angle formed by the 
z~axis and the line connecting the intersection between the parabola
and the equator with the intersection between the parabola
and the outer sphere.

For the dust, we used the mixture of \citet{Mathis77}. The values of the
opacities are obtained using a dust to gas ratio by mass of 
$\sim 10^{-2}$. The optical constants and albedos for this mixture were 
chosen from \citet{DraineLee84} and \citet{Draine85}. The values used for 
the opacity \kkext\ are 2.0, 3.8 and \opac{6.5}, and the values for 
the albedo $\omega_\lambda$\ are 0.22, 
0.34 and 0.45, in the K, H and J bands respectively. The phase function
that describes the scattering is within the Rayleigh approximation
in the K band, and becomes gradually forward throwing towards the J band.

\section[Grid of Models]{Grid of models}
\label{GridSection}

In this section, a grid of models is presented (see Table~\ref{GridTable}) 
to illustrate how variations on the input parameters affect the morphology 
of the model images. All models shown have an outer radius \rout~=~4700~AU, 
and are assumed to be located at a distance of 1~kpc. Since we aim to
compare the model predictions with typical speckle observations in
4m-class telescopes, synthetic images are generated with a size of 
\fovp{128}{128}\ pixels at a pixel scale of 0\farcs06. Our
choice of \rout\ keeps the outer boundary of the models outside the
field of view of the synthetic images. The images were convolved with
a gaussian with a FWHM=0\farcs2, which is the typical resolution
achieved with speckle imaging in 4m-class telescopes (see Paper~I).

The input parameters are varied with respect to a fiducial model 
(K01, in Table~\ref{GridTable}). The fiducial model consists of an 
envelope with a centrifugal radius of 50~AU and a mass infall rate 
\macc{1.11}{-4} (note that \mac $\sim 10^{-8}\ - 10^{-5}$ \macco\ are 
typically inferred for low mass YSOs ; 
\citealp[e.g. ][]{Kenyon93,Lucas97,Whitney97}). The mass infall rates
used in our grid are consistent with envelope models in massive stars 
\citep{WolfireCasinelli86,Maeder02}. This mass infall rates, though,
would be smaller for models with a larger outer radius.  The model also has a
geometrically thin but optically thick disc ($\alpha = 3/4$ and 
$\beta = 0$) of radius \rd~=~250~AU and 
$\rho_{0}$=$2 \times 10^{-4}$\ kgm$^{-3}$. A conical cavity with a radius at
the equator of 100~AU and an opening angle \opa~=~\degg{20} represents
the effect of the outflow in the envelope. The central source is
assumed to be a \msun{10} star with a radius \rst~=~10~\rsuno. The
number of input
photons in model K01 is \eng{5}{6}. Only 8.7\% of these photons
form part of the output (see column~9 in Table~\ref{GridTable}). 
The other 91.3\% is absorbed either in the disc (9.3\% of the total) 
or in the envelope (79.2\%). There is a small fraction (2.8\%) that 
is absorbed by the star itself after being scattered. The optical
depth along the line of sight of the fiducial model in the K band 
(\od = 8 in Table~\ref{GridTable}) corresponds to an extinction in the
K~band of \akapp~=~8.6~mag. If the extinction law of \citet{He95} is
used, this corresponds to a visual extinction of $\sim$80 magnitudes,
i.e. an optical depth of the 9.7~$\mu$m silicate feature of 
\tausi~$\sim$~4, (where $\mathrm{ \tau_{Si}(9.7 \mu m) = 0.053 A_{v}}$ 
from \citealp{DraineLee84} was used), which is typically observed in 
massive YSOs.


  \begin{table}
      \caption[Grid of Models]{Grid of models}
         \label{GridTable}
	 \begin{scriptsize}
         \begin{tabular}{lcccccccc}
            \hline
	    \hline
	    Mod$^\mathrm{a}$   & \rcav & \rce  & \rd   & \od / $\iota$  & Cav$^\mathrm{b}$ & Dis$^\mathrm{c}$ & \mac   & \fout  \\
            \hline
            \noalign{\smallskip}
	    K01   &  100  &  50   &  250  &  8 / 45  &  C20   &  T   &  1.11  &  0.087 \\ \hline
	    K02   &  100  &  50   &  250  &  6 / 45  &  C20   &  T   &  0.47  &  0.119 \\ 
	    K03   &  100  &  50   &  250  &  7 / 45  &  C20   &  T   &  0.77  &  0.093 \\
	    K04   &  100  &  50   &  250  &  9 / 45  &  C20   &  T   &  1.38  &  0.173 \\
	    K05   &  100  &  50   &  250  &  10 / 45  &  C20   &  T   &  1.44  &  0.066 \\ \hline
	    K06   &  100  &  50   &  250  &  8 / 75  &  C20   &  T   &  0.47  &  0.068 \\ 
	    K07   &  100  &  50   &  250  &  8 / 60  &  C20   &  T   &  0.77  &  0.138 \\
	    K08   &  100  &  50   &  250  &  8 / 25  &  C20   &  T   &  1.80  &  0.350 \\ \hline
	    K09   &  100  &  SPH  &  250  &  8 / 45  &  C20   &  T   &  1.17  &  0.087 \\
	    K10   &  100  &  100  &  250  &  8 / 45  &  C20   &  T   &  1.30  &  0.080 \\
	    K11   &  100  &  150  &  250  &  8 / 45  &  C20   &  T   &  1.47  &  0.077 \\
	    K12   &  100  &  200  &  250  &  8 / 45  &  C20   &  T   &  1.62  &  0.074 \\ \hline
	    K13   &  100  &  50   &  250  &  8 / 45  &  C10   &  T   &  1.06  &  0.058 \\
	    K14   &  100  &  50   &  250  &  8 / 45  &  C30   &  T   &  1.18  &  0.124 \\
	    K15   &  100  &  50   &  250  &  8 / 45  &  C40   &  T   &  1.28  &  0.167 \\
	    K16   &  100  &  50   &  250  &  8 / 45  &  C50   &  T   &  1.41  &  0.214 \\ \hline
	    K17   &  50   &  50   &  250  &  8 / 45  &  C20   &  T   &  0.83  &  0.075 \\
	    K18   &  150  &  50   &  250  &  8 / 45  &  C20   &  T   &  1.39  &  0.094 \\
	    K19   &  200  &  50   &  250  &  8 / 45  &  C20   &  T   &  1.66  &  0.101 \\
	    K20   &  250  &  50   &  250  &  8 / 45  &  C20   &  T   &  1.93  &  0.106 \\ \hline
	    K21   &  100  &  50   &  100  &  8 / 45  &  C20   &  T   &  1.11  &  0.087 \\
	    K22   &  100  &  50   &  150  &  8 / 45  &  C20   &  T   &  1.11  &  0.087 \\
	    K23   &  100  &  50   &  200  &  8 / 45  &  C20   &  T   &  1.11  &  0.087 \\
	    K24   &  100  &  50   &  500  &  8 / 45  &  C20   &  T   &  1.11  &  0.087 \\
	    K25   &  100  &  50   &  250  &  8 / 45  &  C20   &  F   &  1.11  &  0.087 \\ \hline
	    K26   &  100  &  50   &  250  &  8 / 45  &  P20   &  T   &  1.58  &  0.150 \\ \hline
	    J01   &  100  &  50   &  250  &  25 / 45  &  C20   &  T   &  1.11  &  0.097 \\
	    H01   &  100  &  50   &  250  &  15 / 45  &  C20   &  T   &  1.11  &  0.055 \\
	    HH01  &  100  &  50   &  250  &  12 / 45  &  C20   &  T   &  1.11  &  0.063 \\ 
            \hline
	    \noalign{\smallskip}
         \end{tabular}		  
	 \end{scriptsize}
      \begin{list}{}{}
      \item[$^{\mathrm{a}}$] Model name, composed of the filter used 
	to generate the image, and a serial number. 
      \item[$^{\mathrm{b}}$] Type of cavity; (P)arabolical or
	(C)onical, and opening angle (in degrees).
      \item[$^{\mathrm{c}}$] Type of disc; (F)lared or (T)hin.
      \item[] \rcav, \rce\ and \rd\ are the cavity, centrifugal and disc
	radius in AU. \od\ represents the optical depth along the line 
	of sight at the wavelenght of the model. The inclination of
	the line of sight in degrees with respect to the system axis
	is represented by $\iota$. \mac\ is the mass 
	accretion rate in units of $10^{-4}\ $\macco. \fout\ is the 
	ratio between the total number of output photons (i.e. the sum
	of scattered and direct photons from the star) in all
	directions and the number of input photons at the wavelenght
	of the model.
      \end{list}
   \end{table}

\subsection[Morphology]{Morphology}

  \begin{figure}
   \centering
   \resizebox{\hsize}{!}{\includegraphics{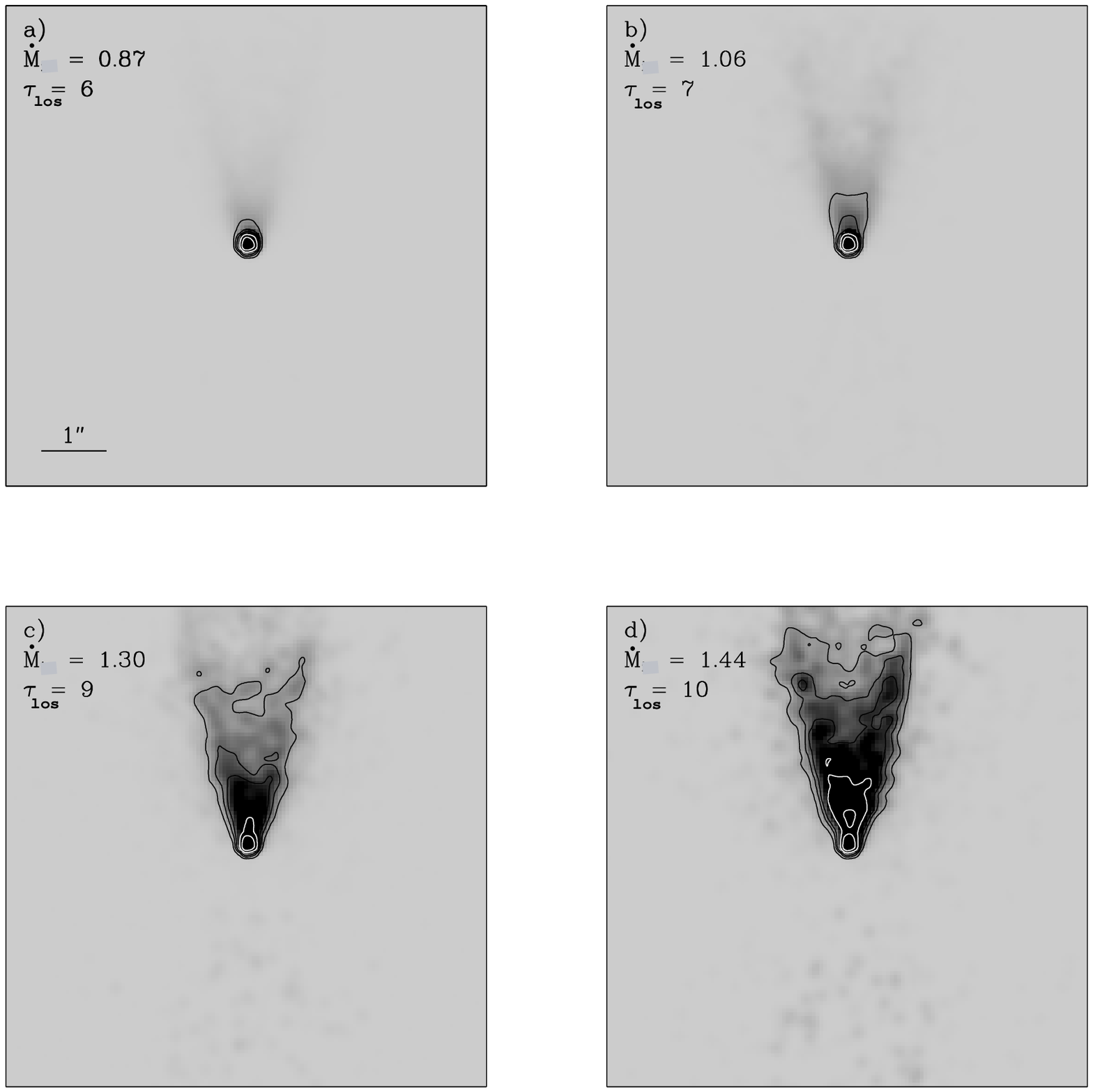}}
   \caption[]{K~band images for models in which the infall rate has
     been varied with respect to the fiducial model (K01 in
     Table~\ref{GridTable}, see Fig.~\ref{ModelFiduFig}c). {\bf a)} Model 
     K02 with a \macc{0.87}{-4}. {\bf b)} Model K03 (\macc{1.06}{-4}).
     {\bf c)} Model K04 (\macc{1.30}{-4}). 
     {\bf d)} Model K05 (\macc{1.44}{-4}). The rest of the parameters 
     remain unchanged with respect to model~K01. In the four panels, the
     line of sight forms an angle of \degg{45}\ with respect to the 
     cavity axis. The 1$''$\ bar represents a length of 1000~AU, and
     the resolution used is 0\farcs2. The greyscale varies from -5\%
     (the lightest) to 20\% of the maximum brightness (the
     darkest). The contours are at 5, 10, 15, 30 and 50\% of the
     maximum brightness. These contour levels have been chosen for
     comparison with typical near-IR speckle imaging in 4m-class 
     telescopes (see Paper~I). \od\ represents the optical
     depth along the line of sight. These and the other model images
     shown in this section are normalised to the brightness peak,
     unless otherwise stated.}
   \label{ModelOpacityFig}
   \end{figure}

In this section, we investigate which parameters of the circumstellar
density distribution can be constrained from the observed morphology
of the nebula.
Fig.~\ref{ModelOpacityFig} shows the effect produced in the synthetic 
images by varying the overall optical depth in the envelope. This is
done by changing the density scaling
through the mass infall rate (\mac) in the Ulrich formula. At
low optical depths (Fig.~\ref{ModelOpacityFig}a), there is
less dust available to scatter the light, and also to absorb the
direct light from the star. Therefore, the nebula becomes relatively 
fainter than at higher optical depths, and the only contribution to
the image at a 5\% level is the central star. However, at higher
optical depths, the nebula becomes much more extended, since the 
central star becomes highly obscured and there is more dust available 
for scattering. Hence, only a change of 50\% in the mass infall rate
has a dramatic change on the appearance of the system, due mainly to
the exponential dependence on the line of sight optical depth of the
central star brightness. Therefore, from an observational point of
view, the line of sight optical depth is the most important parameter.

   \begin{figure}
   \centering
   \resizebox{\hsize}{!}{\includegraphics{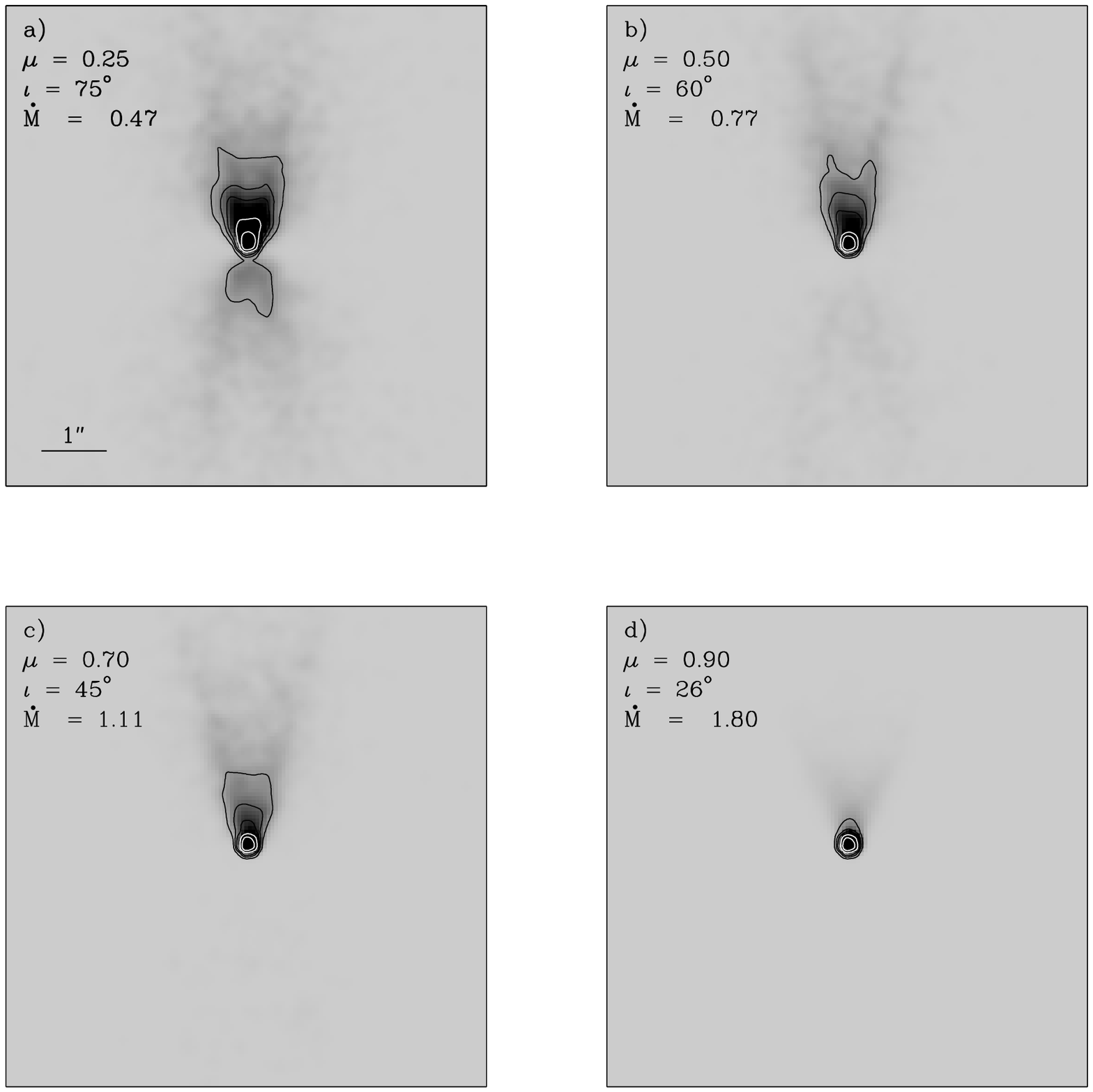}}
   \caption[]{Synthetic K~band images of reflection nebulae for the
     models at different inclination angles ($\iota$). The mass infall 
     rate has been adjusted to yield the same optical depth along the 
     line of sight (\od = 8) at all four inclination. \mac\ is
     expressed in units of $10^{-4}\ $\macco\ and $\iota$\ in
     degrees. Panel~{\bf a)} 
     corresponds with a near-edge-on model (K06 in Table~\ref{GridTable}). 
     The image in panel~{\bf b)} corresponds with an intermediate
     inclination (model K07). Panel~{\bf c)} shows the fiducial model
     (K01), which is seen at an inclination angle of \degg{45}. {\bf d)} 
     Model with a near-pole-on inclination (K08). The contours and the
     greyscale in all panels are defined as in Fig.~\ref{ModelOpacityFig}.}
   \label{ModelFiduFig}
   \end{figure}

Figure~\ref{ModelFiduFig} shows the K~band images for different 
inclination angles of the line of sight with respect to the system 
axis (i.e. cavity axis). The overall density scaling has ben adjusted in
each of the four models to yield the same optical depth along the line
of sight as the fiducial model (\od~=~8). For views near edge-on
(panel a in Fig.~\ref{ModelFiduFig}), the nebula is clearly 
bipolar at the 5\% level. The receding lobe appears less
bright than the approaching lobe. In this case (model K06 in 
Table~\ref{GridTable}), a mass infall rate 2.3 times lower than for the fiducial
model was used. Even at this near-edge-on inclination, it is possible
to see the central star due to the low overall density
scaling. At intermediate inclinations (panels b and c), the receding
lobe is not detected any more at the 5\% level. At low inclinations,
(panel d) a faint monopolar nebula can still be seen. 

   \begin{figure}
   \centering
   \resizebox{\hsize}{!}{\includegraphics{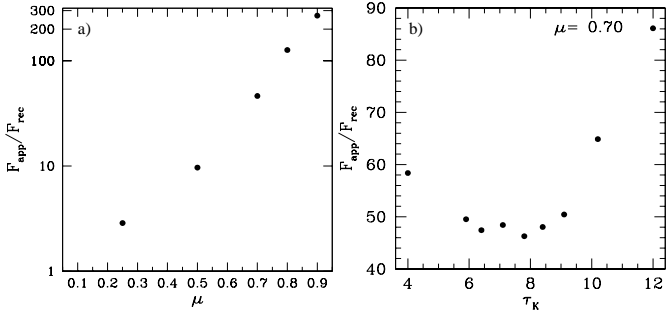}}
   \caption[]{{\bf a} Variation of the approaching to receding lobe flux ratio
     with system inclination angle, where the line of sight optical depth is
     kept constant. The system inclination angle varies from near-edge-on
     ($\mu = 0.1$) to near-face-on ($\mu = 0.9$). {\bf b} Variation of
     the approaching to receding lobe flux ratio with overall opacity.}
   \label{UpDownFig}
   \end{figure}

The contrast between the approaching and receding nebula (\fapprec)  
is a useful quantity to estimate the system inclination angle. 
Fig.~\ref{UpDownFig}a shows the variation of the \fapprec\ with the 
inclination angle. Each point in the plot represents a model whose
overall density scaling has been adjusted to yield an optical depth
along the line of sight of 8 at its corresponding inclination. The
flux ratio has been calculated using aperture photometry with an
aperture radius of $1''$\ in the synthetic images. The aperture on the approaching 
lobe was centered at 0\farcs8 from the star along the cavity axis
and it includes the star itself. The aperture on the receding lobe was 
centered at 1\farcs2 from the star also along the cavity axis and
it does not include the star. This avoids sensitivity to the actual
location of the apertures. For near-edge-on views (small $\mu$'s), 
the approaching and receding lobe have roughly the same brightness. 
Therefore, the \fapprec\ is nearly~1. As $\mu$\ increases, the approaching
lobe becomes relatively brighter, and hence the value of \fapprec\
increases. The figure also shows that the ratio \fapprec\ becomes 100 
for values of $\mu$\ in the range between 0.70 and 0.80 (i.e for 
inclination angles in the range \degg{35} and \degg{45}). Hence, a
1\% upper limit in the detection of the receding lobe indicates
that the system is seen under an inclination angle $\le 45^{\circ}$.

Figure~\ref{UpDownFig}b shows the change of \fapprec\ due to variations 
in the optical depth (i.e. variations in the \mac) for an inclination angle 
of \degg{45}. At very low optical depths (\od~$\la 5$) the star dominates 
the flux. As the opacity increases ($6 \la$~\od~$\la 9$), the star is 
increasingly obscured, and the counter-lobe starts to show up. At larger 
optical depths (\od~$\ga 10$), the approaching lobe starts to dominate 
the emission, while the receding lobe hardly 
changes. Hence, the ratio \fapprec\ increases again.
The contrast between the nebula lobes is less sensitive to changes in
the mass infall rate than to changes in the inclination angle. Hence, 
in principle, the approaching to recceding lobe flux ratio can be used 
to constrain the inclination angle.

   \begin{figure}
   \centering
   \resizebox{\hsize}{!}{\includegraphics{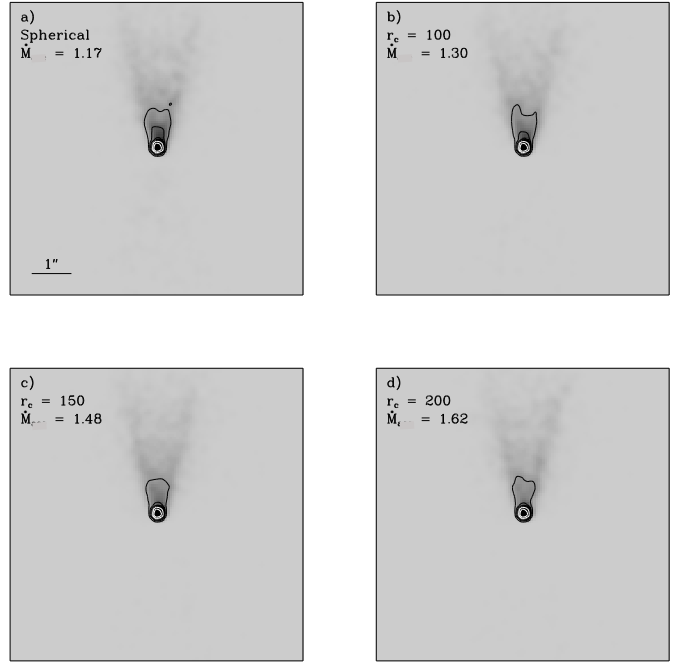}}
   \caption[]{K~band images resulting from models with different
     centrifugal radius seen at an inclination angle of \degg{45}. The
     mass accretion rate for each model was set to keep the \od~=~8
     along the line of sight at \degg{45} in the K~band. The rest of
     the parameters remain unchanged with respect to the fiducial
     model. In all panels, the value of \rcen\ is expressed in AU and 
     \mac\ is expressed in units of $10^{-4}\ $\macco. {\bf a)} Model~K09
     in Table~\ref{GridTable}, which represents a spherical envelope (i.e. the 
     limiting case of a negligible centrifugal radius). {\bf b)} Model~K10, 
     which has a \rcen\ = 100~AU. {\bf c)} Model K11, which has a 
     \rcen\ = 150~AU. {\bf d)} Model K12 with a \rcen\ = 200~AU. The
     contours and the greyscale in all four panels are defined as in 
     Fig.~\ref{ModelOpacityFig}.}
   \label{ModelRcenFig}
   \end{figure}

In Fig.~\ref{ModelRcenFig}, we investigate whether the shape of the
envelope can be derived. The images show models with an increasing
centrifugal radius (i.e. flatteness of the envelope), from panel a) to
panel d). The mass infall rate for each model has again been set such that the
optical depth along a line of sight at \degg{45}\ remains the same as 
for the fiducial model (i.e. \od\ =~8 in the K~band). The fiducial
model, with a centrifugal radius of 50~AU, is shown in 
Fig.~\ref{ModelFiduFig}c. Figure~\ref{ModelRcenFig}a 
represents the case of a spherical density distribution.
The nebula becomes less bright as the centrifugal radius 
increases (panels b, c, and d in Fig.~\ref{ModelRcenFig}). 
Since the material is predominantly concentrated on the equatorial plane 
and there is less dust available in the polar regions of the envelope,
where a large fraction of the scattered light is generated. However,
this is a very subtle change compared to that for the line of sight
optical depth o inclination angle. Hence, the reflection nebula tells
us little about the degree of flattening of the envelope. 

\begin{figure}
   \centering
   \resizebox{\hsize}{!}{\includegraphics{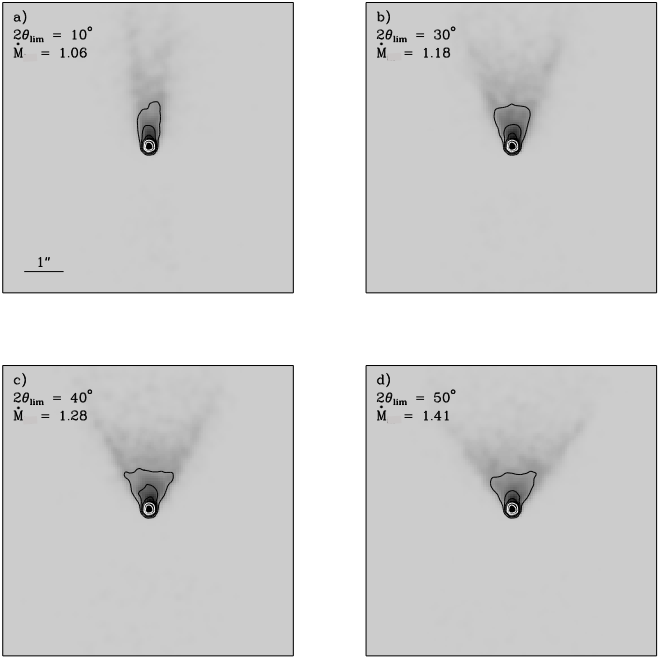}}
   \caption[]{K~band images corresponding to models with different 
     cavity opening angles
     seen at an inclination angle of \degg{45}. The \mac\ for each model 
     (which is shown in units of $10^{-4}\ $\macco) has been set to
     yield a K~band optical depth of 8 along the line of sight. The 
     conical cavity opening angle increases from panel 
     {\bf a)} (\degg{10}) to panel {\bf d)} (\degg{50}) (i.e. models
     from K13 to K16 in Table~\ref{GridTable}). The  greyscale and 
     contours are as in Fig.~\ref{ModelOpacityFig}.}
   \label{ModelAngleFig}
   \end{figure}

Figure~\ref{ModelAngleFig} shows the effect of varying the cavity 
opening angle ($10^{\circ} \le 2\theta_{\mathrm{lim}} \le 50^{\circ}$) 
on the synthetic nebulae.  The mass infall rate was increased with the 
cavity angle to keep the same optical depth along the line of sight in 
all images. Unsurprisingly, the nebula opening angle 
appears to be larger for models with a wide-angled cavity. 
The significant changes on the nebula shape occur for the external contours, 
while the inner contours remain nearly unchanged. This is consistent
with the fact that variations in the cavity opening angle will affect 
more the external regions of the envelope than the regions close to 
the cavity base. At wider cavity opening angles, more stellar photons 
can escape the system without being scattered. This is shown in
column~9 of Table~\ref{GridTable}, where the fraction of photons 
(scattered and stellar) that leaves the system (\fout) is listed. 

\begin{figure}
   \centering
   \resizebox{\hsize}{!}{\includegraphics{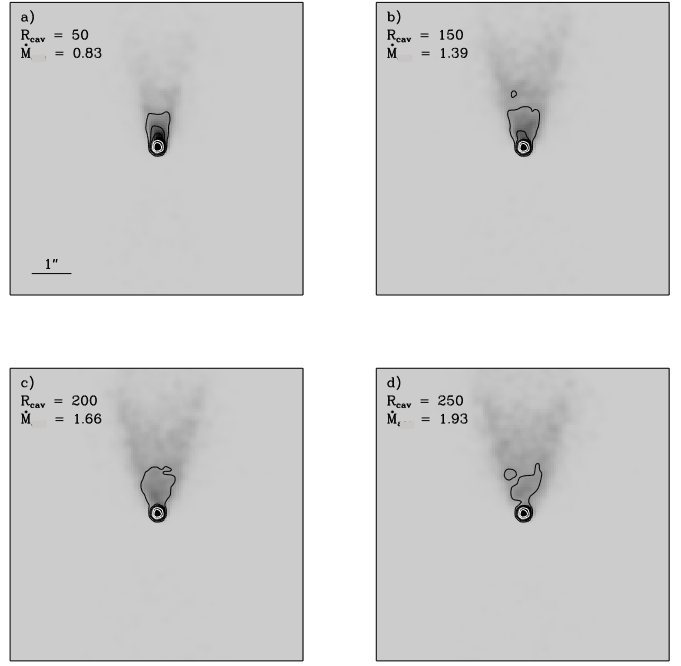}}
   \caption[]{Variation of the nebula shape with the cavity radius at
     the equator (models K17 to K20 in Table~\ref{GridTable}). The
     mass infall rate was adjusted to yield the same optical depth
     along the line of sight in all four images. The values for \rcav\ 
     are expressed in AU. All panels represent a line of sight at 
     \degg{45}. The greyscale and
     contours are as in Fig.~\ref{ModelOpacityFig}.}
   \label{ModelRcavFig}
   \end{figure}

The result of varying the cavity radius at the equator from 50 to
250~AU is shown in Fig.~\ref{ModelRcavFig}. A radius of 50~AU
corresponds approximately to the sublimation radius for dust in OB
stars. Radiation pressure or wind interactions could increase the
size of the hole. The nebula appears fainter and slightly wider for 
larger values of the \rcav. In this case, 
the differences between nebulae can also be observed in the innermost 
contours. This is caused by the fact that changes in \rcav\ will have 
a stronger impact in the regions of the envelope closer to the equator 
than in the outer regions. The total number of output photons increases 
by a factor of 1.4 from the model with \rcav\ = 50~AU to the model
with \rcav\ = 250~AU (see column~9 in Table~\ref{GridTable}). For
larger values of \rcav, similar number of photons are scattered of a 
larger surface. Hence, the surface bightness of the nebula decreases
from Fig.~\ref{ModelRcavFig}a~to~d. 

\begin{figure}
   \centering
   \resizebox{\hsize}{!}{\includegraphics{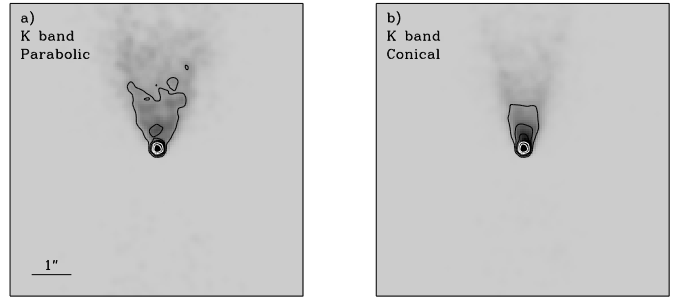}}
   \caption[]{{\bf a)} Model with a parabolic cavity (K26 in 
     Table~\ref{GridTable}) compared with the fiducial model at an
     inclination of \degg{45}. The cavity opening angle is the same as
     for the conical cavity in the fiducial model (\opa~=~\degg{20}). 
     {\bf b)} Fiducial model at an inclination of \degg{45}. This is the
     same image as shown in Fig.~\ref{ModelFiduFig}c.  The greyscale
     and contours are as in Fig.~\ref{ModelOpacityFig}.  The density
     scaling is a factor of 1.4 larger in the model with the parabolic
     cavity than in the conical cavity model.}
   \label{ModelParaFig}
   \end{figure}

We now investigate the effect of the detailed shape of the base of the
cavity. A parabolic cavity might be expected if the central wind is
initially equatorial, whilst a conical cavity would arise from a
jet-driven flow in low mass stars \citep{Bachiller95}. 
Figure~\ref{ModelParaFig}a shows the K band image for a model with a 
parabolic cavity compared with a conical nebula in
Fig.~\ref{ModelParaFig}b. The other parameters have
the same values as for the fiducial model, except the overall opacity
scaling, which has been enhanced by a factor of 1.4 to yield the same
optical depth along the line of sight as in the fiducial model. The
parabolic shape is clearly seen in the resulting nebula, as expected 
(Fig.~\ref{ModelFiduFig}). A larger fraction of photons escape from the 
parabolic model than from the conical model because the parabolic
cavity is broader near the star than the conical cavity. The concave shape 
of the cavity walls, favours the scattering in the outer regions of the 
envelope, which also contributes to make the parabolic nebula more extended.

All models shown up to this point included an optically thick, and 
geometrically thin flat disc of radius \rd~=~250~AU. To investigate 
any dependence of the model images on the disc radius, we have
generated models with the same parameters as the fiducial model (K01, 
Fig.~\ref{ModelFiduFig}a) but with varying disc radius 
($\mathrm{100~AU \le \mathrm{r_d} \le 500~AU}$). No significant
differences were found between the images resulting from these models
and the fiducial model. We also investigated whether the introduction of a
flaring disc may change the morphology of the reflection
nebulae. Model K25 in Table~\ref{GridTable} has the same parameters as
the fiducial model except for a flaring in the inner disc. 
The values used for $\alpha$\ and $\beta$\ in the disc equation 
(Eq.~\ref{eq48}) were 15/8 and 9/8 respectively (i.e. a Keplerian disc). 
No relevant differences were appreciated between these images and the 
fiducial model. The reason is that the disc flaring angle for the
value of $\beta = 9/8$\ is small compared with the line of sight
inclination angle. The envelope density dominates the disc density for all
inclinations $\le 77^{\circ}$\ ($\mu \ge 0.22$).

\subsection[Dependence with wavelength]{Dependence with wavelength}
\label{HKSubSection}

   \begin{figure}
   \centering
   \resizebox{\hsize}{!}{\includegraphics{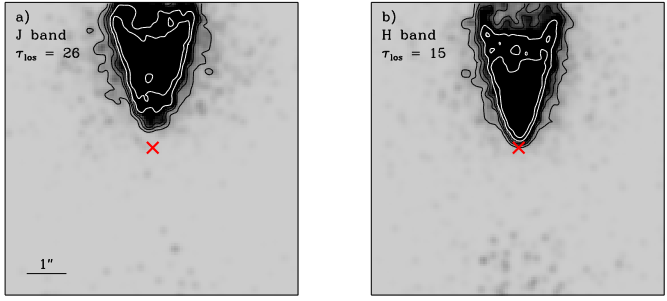}}
   \caption[]{Multi-band models. The two panels represent the fiducial
     model in the J~and H~bands, seen at an inclination angle of
     \degg{45}. In Table~\ref{GridTable} they are labeled as J01
     (panel~a), H01 (panel~b). $\tau$\ is the optical depth at 
     an inclination of \degg{45}. The greyscale and the contours are 
     the same as in Fig.~\ref{ModelOpacityFig}, and the grey cross 
     indicates the position of the star.}
   \label{HKModelFig}
   \end{figure}

Figures~\ref{HKModelFig}a, \ref{HKModelFig}b and~
Fig.~\ref{ModelFiduFig}c show the variation of the synthetic images 
with the wavelength (J,~H and~K band respectively). The geometry and 
density distribution for the J and H~band models are the same as for the 
fiducial model. However, the opacities, the albedos and the scattering 
matrix change with wavelength. The opacities correspond with an extinction
law ($\mathrm{\kappa^{ext}_{\lambda} \propto \lambda^{-\gamma}}$) with 
an exponent $\gamma = 2.3$. The number of photons in the input
spectrum at~H ($\mathcal{N}_{\mathrm{H}}^\mathrm{in}$) is a factor of 2.00 times the
number of photons at K ($\mathcal{N}_{\mathrm{K}}^\mathrm{in}$), and the number of photons
at~J ($\mathcal{N}_{\mathrm{J}}^\mathrm{in}$) is 2.45 times the number of photons at
K. The number of input photons at each wavelength was calculated using 
the model stellar atmospheres from \citet{Kurucz79}. The spectral
energy distribution (SED) of model OB main sequence stars
were integrated in the J, H and K bands using the transmission
profiles of the filters J98, H98 and K98 at UKIRT. The ratio between
the number of photons emitted at two given bands approaches
asymptotically to a constant value for earlier spectral types. This
limit corresponds with a slope in the stellar SED of -2.4 
($\mathcal{N}_\lambda^\mathrm{in} \propto \lambda^{-2.4}$), which is shallower than the
theoretical Rayleigh-Jeans limit ($\mathcal{N}_\lambda^\mathrm{in} \propto \lambda^{-3}$). 
This shallower SED is a better representation of the colours
of OB main sequence stars than the Rayleigh-Jeans approximation (see
the UKIRT web-page: {\tt http://www.jach.hawaii.edu/JACpublic/UKIRT/}). 
Figures~\ref{HKModelFig}a, \ref{HKModelFig}b 
and~\ref{ModelFiduFig}c  show that the nebula appears
more extended at short wavelengths, while the star becomes totally obscured,
due to the increase in the opacity. It can also be seen that the separation 
between the star and the nebula apex decreases towards longer wavelengths,
i.e. it is possible to probe the circumstellar density distribution closer 
to the star at longer wavelengths.

We now address the question of how the colours of the synthetic images
vary with the different model parameters. 
The \hk\ colour for the models can be estimated using the following expression,

\begin{equation}
\label{eq434}
    \mathrm{(H - K)_{mod}} = \mathrm{-2.5\log\left(\frac{\mathcal{N}^\mathrm{out}_{H}}{\mathcal{N}^\mathrm{out}_{K}}\frac{\lambda_{K}}{\lambda_{H}}\frac{f_{K}^{0}}{f_{H}^{0}}\right)} \nonumber \\
\end{equation}


where the $\mathcal{N}$'s represent the number of photons  within a
certain aperture
on each band. The $\lambda$'s are the central wavelengths of the filters, and
the $\mathrm{f^{0}}$'s are the zero magnitude fluxes. The values used for 
$\lambda$\ are 1.65 and \mum{2.20}\ at~H and~K respectively, and
\mbox{$1.12\times10^{-9}$~Wm$^{-2}\mu$m$^{-1}$}\ and 
\mbox{$4.07\times10^{-10}$~Wm$^{-2}\mu$m$^{-1}$}\ are the values for 
\floh\ and \flok\ respectively (UKIRT web-page:
{\tt http://www.jach.hawaii.edu/JACpublic/UKIRT/}). For simplicity, 
the colous shown here assume that the foreground extinction is
negligible. Note that the presence of unknown foreground extinction is
an added complication when models are compared with observations.

   \begin{figure}
   \centering
   \resizebox{\hsize}{!}{\includegraphics{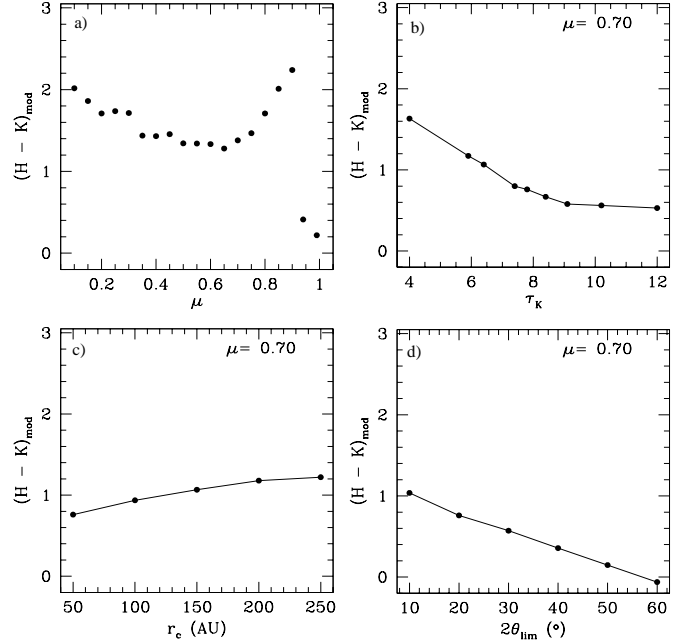}}
   \caption[]{{\bf a)} Variation of the model \hk\ colour
     (Eq.~\ref{eq434}) with the line of sight inclination 
     angle for the fiducial model. {\bf b) - d)} The inclination angle 
     is fixed 
     at \degg{45}\ and the overall opacity, centrifugal radius and cavity 
     opening angle are varied. In panels c) and d) the mass infall rate was 
     also changed to keep the opacity along the line of sight
     constant.}
   \label{HKFig}
   \end{figure}

Figure~\ref{HKFig}a shows the change in the \hk\ colour with the line of 
sight inclination angle for the fiducial model. An aperture radius of 
1$''$\ centered at the image centre was used. The \hk\ colour at all
inclinations, except for the very close to pole-on, is clearly redder
than the \hk\ colour of the input spectrum ((\hk)$_\mathrm{inp}$=0.04). At
edge-on inclinations ($\mu \sim 0$), no star is seen in 
either band, and the nebula at K is less extincted than the nebula at
H. At intermediate inclinations, the nebula at H becomes brighter
because the stellar light passes through a less dense part of the envelope. The
star begins to appear at K but is not yet seen at H. The overall effect
is that the \hk\ colour becomes bluer. For $\mu > 0.65$\ the 
star brightens quickly at K, dominating the flux in this 
band, while the star just begins to appear at H. Hence, the 
\hk\ colour becomes redder. At face-on views (i.e. $\mu \ge 0.94$) the
\hk\ colour tends to the (\hk)$_\mathrm{inp}$ because the
star is now seen directly through the cavity. The behaviour of the
\hk\ colour with the inclination angle in our simulations is different
to that found by \citet{Kenyon93} and \citet{Whitney97}. Their models
become bluer at high inclination angles, while ours become redder due
to the higher envelope optical depth (i.e. higher mass infall rate)
and due to the fact that, at high inclinations,  no star is seen in
the H band. 

Figure~\ref{HKFig}b illustrates how the \hk\ colour varies with the
overall opacity at a fixed inclination angle of \degg{45}. At large 
opacities, the \hk\ colour is bluer because the reflection nebula
becomes relatively brighter than the star. In Fig.~\ref{HKFig}c, we
represent models with different centrifugal radii seen at a fixed 
inclination of \degg{45}. The mass accretion rate was adjusted so that 
all of them have an optical depth along the line of sight of 8. In the models
with a larger centrifugal radius (i.e. a flatter density distribution)
there is less dust available in the outer parts of the envelope, since
the dust is mainly concentrated at the equator. This density enhancement in
the equatorial region allows more direct light from the central star
to escape (predominantly in the K band due to the low opacity) than in
models with a small \rcen. The consequence is that the \hk\ colour becomes
slightly redder for models with large \rcen. 

Figure~\ref{HKFig}d shows variations of the \hk\ colour with the
cavity opening angle. All models represented in this plot are seen at
an inclination of \degg{45}, and the overall density scaling was
adjusted to yield an \od~=~8. The general trend is that at wide cavity 
opening angles the \hk\ colour becomes bluer because more scattered
photons in the outer parts of the envelope (at H, not at K) can escape 
from the system. This general trend is favoured by the fact that the direct
light from the star is equally extincted at all cavity opening angles 
shown in Fig.~\ref{HKFig}d. The figure also illustrates that the \hk\
colour for models with an \opa~$>$~\degg{50} becomes even bluer than
the \hk\ colour of the input spectrum. In summary, the four panels in
Fig.~\ref{HKFig} show that the change in the \hk\ colour is not
higher than 1 magnitude within the ranges of the parameter space 
investigated. Hence, the integrated colour is not particularly
sensitive to the density distribution, and is unlikely to yield unique
solutions for the model parameters.

\subsection[Dust in the Cavity]{Dust in the cavity}
\label{DustCavitySubSection}

Up to now, we have assumed that the cavity evacuated by the outflow is
empty. However, in a more realistic situation, one would expect that
some dust may remain inside the cavity. We estimated the expected
\avio\ due to dust within the cavity from the typical \molh\ column
densities in massive outflows as follows. \citet{RidgeMoore01} show that the
the integrated CO~(J=1-2) intensity in molecular outflows from 11 massive 
YSOs is, on average, $\sim$\,200\,K\kmsc. If we use the expression
$\mathrm{N(H_2)\,(cm^{-2})\,=\,3\times10^{20}\int{T_{mb}dv}}$, where 
$\mathrm{\int{T_{mb}dv}}$\ represents the integrated CO~(J=1-2) intensity 
\citep{Osterloh97,Henning00}, typical $\mathrm{H_2}$\ column densities
in the outflow of $6\times10^{22}$\,cm$^{-2}$\ are obtained. This
yields typical visual extinctions of $\sim\,3.5$\,mag, if the
expression $\mathrm{A_v/N(H_2)\,(cm^{-2})\,=\,5.88\times10^{-23}}$\
from \citet{Bohlin78} is used. This implies an A$_\mathrm{K}\sim\,0.4$\,mag
which is about 200 times smaller than the A$_\mathrm{K}$\,=8, used in
our fiducial model.

To see the effect that dust in the cavity has in our simulations, we have
run three models with the same parameters as our fiducial model but with
an \akapp\ along the cavity axis of 0.1, 1 and 10 mags. A dusty density
distribution within the cavity given by the expression 
$\mathrm{\rho(z)\,=\,(1\,+\,|z|/R_{cav})^{-2}}$, which is expected
in a disc wind with a constant flow density distribution in a conical
cavity, was chosen. The output from the models with \akapp\,=\,0.1 and 
1 are hardly distinguishable from
the fiducial model.  For \akapp\,=\,10, only a few photons from the
star can scape the system, and the reflection nebula is not seen at
all. However, the latter value of \akapp\ is far larger than our
previous estimate of extinction due to dust in the cavity. Therefore,
for the typical
extinction expected due to dust within the cavity in massive YSOs, no 
noticeable effect on the observable properties of our scattering
models is detected. Note that this result depends on the selected
shape for the 
density distribution within the cavity. If an optically thin uniform
density distribution is chosen, the resulting reflection nebula is 
expected to be more extended than for an empty cavity \citep[see ][]{Lucas96}. 

\section[Model for Mon\,R2\,IRS3~S]{Model for \mon~S}
\label{IRS3ModelSection}

\begin{figure}
   \centering
   \resizebox{\hsize}{!}{\includegraphics{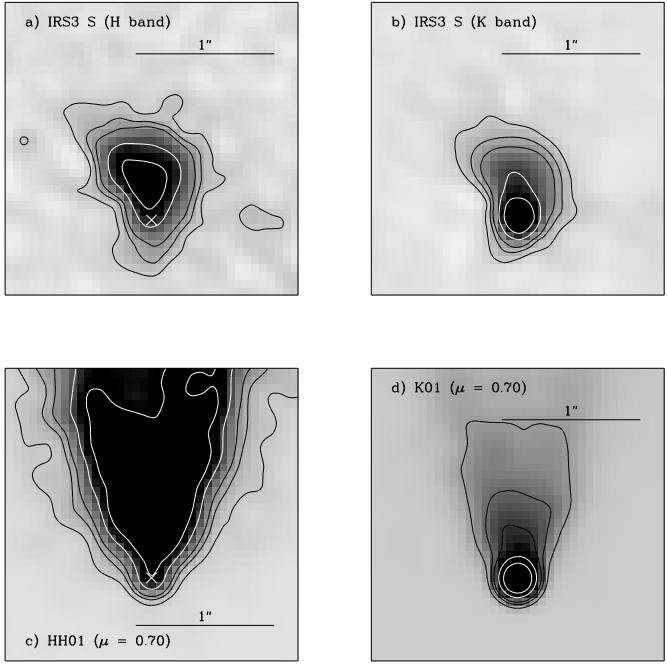}}
   \caption[]{{\bf a)}, {\bf b)} H and K speckle images of
     \mon~S. These images were obtained by rotating the images shown
     in Paper~I by an angle of \degg{198}. The orientation is such
     that the presumed cavity axis is along the vertical 
     axis. The greyscale varies from -5\% (white) to 40\% (black) of
     the brightness peak. The contour levels are at 5, 10, 15, 30 and
     50\% of the peak. The images are normalised to the peak
     brightness. {\bf c)}, {\bf d)} Images for models K01 and HH01
     seen at an inclination angle of \degg{45}. These 
     images are also normalised to their brightness peak. The
     resolution, contours and greyscale are the same as on panels 
     a and b. The cross in the H~band images indicate the location of
     the embedded star.}
   \label{IRS3SModelFig}
   \end{figure}

In Paper I, multi-colour (H and K~band) speckle images of 
a pair of outflow cavities in \mon\ were presented. In this section, 
the data for IRS3~S are used to find observational constraints to the
density distribution using the models presented in the previous
section. In Figs.~\ref{IRS3SModelFig}a and~\ref{IRS3SModelFig}b, we show 
the H~and K~band images of \mon~S shown from Paper~I, with the 
presumed cavity axis oriented along the vertical direction. These
images have a pixel scale of 0\farcs057 and a resolution of 0\farcs19,
which is comparable to the values used the simulations. 
The procedure used to find a good fit model for these data is the following. A
sub-set of models 
that match the morphology of IRS3~S at K are chosen amongst the grid shown 
in Sect.~\ref{GridSection}. A model is searched amongst these that also 
matches the morphology of the source in the H~band. Finally, the \hk\ 
colour of the model is compared with the observations.

Firstly, a reasonable match to the K~band image for IRS3~S was found, 
which is given by the fiducial model used in the grid seen at an inclination 
angle of \degg{45} (Fig.~\ref{ModelFiduFig}c). A zoom into the central 
$2''$\ of this model is shown in Fig.~\ref{IRS3SModelFig}d.
The H~band image for the fiducial model using the opacity  
\kkextH\ = \opac{3.8}\ (c.f. Sect.~\ref{ModelsSection}) appears 
to be too extended (see Fig.~\ref{HKModelFig}b). Other combinations
of inclination angle and mass infall rate were also too extended since
the optical depth at H was always too high. Therefore, a shallower 
opacity law was used in an attempt to achieve a better fit to the IRS3~S 
H~band image. The new value used for \kkext\ at~H was \opac{2.9} which
corresponds with an opacity law ($\kappa_\lambda \propto \lambda^{-\gamma}$)
with $\gamma = 1.3$\ between \mum{1.65}\ and \mum{2.2} (instead of 
the previous value of $\gamma = 2.3$). The exponent 1.3 was chosen to
yield an opacity law slightly flatter than than the $\gamma = 1.7$\
inferred from studies on the interstellar extinction 
in the near-IR \citep[e.g. ][]{He95}. The model using the shallower
opacity law (HH01 in Table~\ref{GridTable}) is shown in 
Fig.~\ref{IRS3SModelFig}. It can be seen that this law still yields 
a rather extended nebula compared with the H~band image of IRS3~S. 
A shallower opacity law than the one given by \citet{DraineLee84} 
has been used by other authors \citep[e.g. ][]{Lucas98} to explain 
the small variation with wavelength of reflection nebulae in 
low mass YSOs. This is possibly due to a different dust composition 
or grain size distribution in the circumstellar matter of YSOs than 
in the interstellar medium. Another possible 
explanation relies on the fact that speckle imaging acts as a spatial 
filter for diffuse extended features of the same order and larger than 
the seeing. Therefore, even if the nebula in the H band was extended for 
a few arcseconds, the reconstructed speckle image would only pick up 
the structure closer to the nebula peak, which is dominated by the 
high spatial frequencies. It may also be the case that the envelope
density distribution is truncated at a distance of $\sim 1000$\ from 
the central source, causing both H and K images to have the same
extent.

Even though a reasonably good fit was found to the K~band speckle image
of IRS3~S, the synthetic nebula shows some differences with respect to the
observed nebula. In particular, the observed nebula departs clearly
from the axi-symmetry, probably due to foreground extinction or the
presence of a clumpy envelope. Besides, the brightness of the synthetic nebula
(Fig.~\ref{IRS3SModelFig}d) drops faster near the star than in the 
case of the observed nebula (Fig.~\ref{IRS3SModelFig}b). Further out, 
at $\sim 1''$\ from the star, the brightness of the synthetic nebula appears 
to drop more smoothly than observed in IRS3~S. The sub-arcsecond
nebula is the inner region of a large reflection nebula (with a diameter of 
$\sim 15''$) that is seen in the near-IR seeing limited images of \mon\ 
\citep[see][]{Aspin90,Yao97}.
The large scale nebula is a halo of reflected light around the whole cloud
but its polarisation pattern indicates enhanced scattering along an
axis that coincides with the sub-arcsecond nebula axis (PA~$\sim$~\degg{198}) . 

   \begin{figure}
   \centering
   \resizebox{\hsize}{!}{\includegraphics{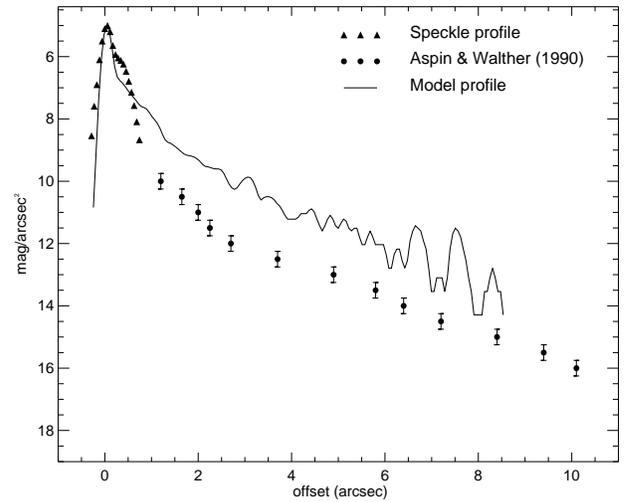}}
   \caption[]{K~band profile along the cavity axis for IRS3~S. The speckle data
     (filled triangles) are obtained from Fig.~\ref{IRS3SModelFig}b. The 
     solid circles represent the data from the contour map shown by
     Aspin \& Walther (1990). The solid line corresponds with an
     extension of the fiducial model`up to a radius of
     \anyexp{9.4}{4}{AU}. The speckle data and the model are
     flux calibrated following the procedure explained in the
     text. The offsets are in arcseconds from the position of the
     star. The errors for the speckle data are smaller than the size of
     the plotting symbol.}
   \label{IRS3AspinFig}
   \end{figure}

To investigate if the model that matches the sub-arcsecond morphology of
IRS3~S fits the outer part of the reflection nebula, the outer radius of 
the fiducial model was doubled while the mass infall rate was slightly 
reduced to \macc{1.03}{-4}\ to keep a K~band optical depth of 8 at an 
inclination of \degg{45}. The profile along the cavity axis for this model 
was compared with the profile for the K~band reconstructed image of IRS3~S 
and with the profile along the same 
direction for the reflection nebula studied by \citet{Aspin90}. This
comparison is shown in Fig.~\ref{IRS3AspinFig}. The data for the large 
scale nebula were taken from the flux-calibrated contour map shown in 
Fig.~3 of \citet{Aspin90}.  The
error bars correspond to half of the contour level separation in their map 
(\magarc{0.25}). The data points for the observed sub-arcsecond nebula were 
calculated from a 0\farcs2 wide line along the cavity 
axis passing through the star in IRS3~S on the normalised speckle image. The 
zero point offset for the magnitude scale was calculated by assigning a K 
magnitude of 6.6 (obtained from the 2MASS survey) to the total number
of counts within an aperture of radius 2\farcs5 that included both 
IRS3~N and~S. The statistical error in the surface brightness is 
\magarc{\la 0.2}.

The profile for the model was calculated from 0\farcs2 wide line along
the cavity axis in the synthetic image normalised to the 
brightness peak. The zero point offset for the magnitude scale was chosen 
such that it yields the same brightness at the peak as the flux-calibrated 
speckle data.

Figure~\ref{IRS3AspinFig} shows that there may be a discrepancy of at least
\magarc{1}\ between the data from Aspin et al. and the speckle data. 
A possible explanation is that speckle imaging filters out any background 
smooth emission, and hence less counts in the speckle image than in the 
seeing limited image of Aspin et al. appear to account for the same flux. 
The profiles plotted in Fig.~\ref{IRS3AspinFig} show that the model 
predicts emission from the outer parts of the nebula (at a distance 
from the star $> 1''$) that is about 2 magnitudes brighter than
it is observed. This may be caused by a drop in the density distribution of 
IRS3 at a radius $\sim 1000$~AU that is not included in the models 
(cf. Figs.~\ref{IRS3SModelFig}a and b). 

\citet{Willner82} estimated an optical depth for the silicate absorption
feature at \mum{9.7} of \tausi~=~4.30, using a $27''$\ aperture
centered at \mon. The same value is also inferred from the ISO spectrum 
of the region (Jackie Keane private comunication). This value can be used 
as an estimate of the total extinction (foreground and through the 
envelope) towards IRS3~S. This \tausi\ leads to an optical extinction of 
\aviapp{82} \citep{DraineLee84} . This corresponds with an extinction
in the K~band \akapp~=~8.6~mag (i.e $\tau_{\mathrm{K}} \sim 7.9$)
using the reddening law of \citet{He95}, which is in good agreement
with the K~band optical depth for the fiducial model 
($\mathrm{\tau_{K} = 8}$).

We found that the \hk\ colour of IRS3~S is redder than the \hk\ colour
of any of the model images. This is probably caused by the fact that
we used the SED of an OB main sequence star to define the input
spectrum. However, it is very likely that the illuminating source has
an excess in the K band due to the presence of a possible accretion
disc, and/or hot dust near the star. The inclusion of an excess of K
band photons in the input spectrum may solve the discrepancy between
the observed and modeled colours. High resolution photometry at longer
wavelengths is required to constrain the input spectrum in the simulations. 

\section[Conclusions]{Conclusions}
\label{ConclusionsSection}

Radiative transfer Monte Carlo simulations have been used to
investigate the density distribution in massive YSOs at scales where
the outflow is generated. The assumed density distribution consists of
a central massive star within a flattened dusty envelope, with a
cavity and an inner optically thick disc. It is found that envelopes with 
density distributions corresponding to typical mass infall 
rates of $\sim 10^{-4} $\macco\ seen at an inclination angle of 
$\sim $\degg{45} approximately reproduce the morphology and extension of 
the sub-arcsecond nebulae observed in massive YSOs. The inclination
angle can be constrained by the measurement of the contrast between
the approaching and the receding nebular lobe, although observations
with a high dynamic range are required (e.g. adaptive optics). The
cavity opening angle is well constrained by the nebula opening angle. 
The simulations indicate possibly some constraints on cavity shape and 
radius at the equator, which could have implications for the
initial angle of the outflow (e.g. jet, wide-angled, equatorial). However,
higher resolution than provided by speckle imaging in 4m-class
telescopes is needed to achieve better constraints of these two 
quantities. The models do not provide significant constraints on 
the flattening in the envelope or the size of the equatorial disc,
which require direct observations with millimetre interferometry.

The Monte Carlo code was also applied to the near-IR sub-arcsecond 
reflection nebula seen in \mon~S. An envelope
with a mass infall rate of $10^{-4}$~\macco\ that includes a conical 
cavity with an opening angle of \degg{20}\ seen at an inclination 
angle of \degg{45}\ provides a reasonable match for the K~band image. 
However, no set of input parameters was found that reproduces both
the~H and~K band images of IRS3~S. An opacity law
with an exponent $\gamma = 2.3$\ \citep{DraineLee84,Draine85} yields 
H band nebulae that are too extended with respect to the observations. 
This would also be the case for the observed interstellar extinction
law ($\gamma = 1.7$; \citealp{He95}).
A shallower opacity law ($\gamma =1.3$) yields a better match to 
observed H~band nebula, although still too extended. This
indicates that the dust in the circumstellar envelope of massive YSOs
may have a rather different optical properties to the dust that forms 
part of the interstellar medium. However, a truncated density
distribution could also explain the data.

Overall, this work shows that future high resolution
($\sim$~0\farcs05) high dynamic range ($> 100$) near-IR imaging has
the potential to constrain the inclination angle and shape of the base
of the outflow cavity. In turn, this could test hydrodynamic models of
the interplay between the infall and outflow in massive YSOs.

\begin{acknowledgements}
  CA would like to thank to the Physics and Astronomy Department
  at Leeds University for their support. CA is also deeply
  grateful to Kapteyn Astronomical Institute for allowing him to use their
  facilities during the realization of this work. We would like to
  thank B. Whitney, the referee of this work, for her comments and 
  suggestions. 
\end{acknowledgements}

\bibliographystyle{aa}
\bibliography{./Alva0318.bib}

\end{document}